\documentclass[a4paper,11pt,]{article}
\usepackage[english]{babel}
\usepackage{geometry}
\usepackage[cp1250]{inputenc}
\usepackage[centertags]{amsmath}
\usepackage{amsfonts}
\usepackage{amssymb}
\usepackage{amsthm}
\usepackage{graphicx}
\usepackage{wrapfig}
\usepackage{epsfig}
\usepackage{multicol,latexsym,a4wide}
\usepackage{lscape}
\usepackage{epstopdf}
\usepackage{lscape} 
\usepackage{pifont}

\usepackage[T1]{fontenc}      

\theoremstyle{definition}

\theoremstyle{definition}

\theoremstyle{remark}

\numberwithin{equation}{section}
\numberwithin{thm}{section}
\numberwithin{lemma}{section}
\numberwithin{cor}{section}
\numberwithin{dfn}{section}
\numberwithin{post}{section}
\numberwithin{rmk}{section}

\begin{document}

\author{M. \v{C}ERMAK}
\title{Electro-vacuum cylindrical stationary solutions in general relativity}

\maketitle

\begin{abstract}
The paper presents a systematical study of stationary (rotating) cylindrical space-times of a Weyl form, solving $D=4$ Einstein-Maxwell equations. The equations are integrated, several classes of exact solutions are obtained by restricting the range of free  parameters, and sub-cases corresponding to already known solutions are identified.
\section*{Keywords} Rotating, electro-vacuum, cylindrical, general relativity, exact solutions, Einstein-Maxwell equations.
\end{abstract}

\section{Introduction}

Searching for exact solutions to Einstein equations is a difficult task in general. The equation of motion in $D=4$ are ten partial differential equations with four variables. These equations can be simplified by imposing symmetry.

One of the simplest solutions is the Schwarzschild solution \cite{Landau}. The Schwarzschild space-time is a spherically symmetric solution to vacuum Einstein equations and describes space-time generated by a massive spherical object. Generalization of the Schwarzschild space-time was found by Reissner and Nordström \cite{Stephani2003}. They found a spherically symmetric electro-vacuum solution to Einstein equations. The Reissner-Nordström solution describes space-time around a massive charged spherical object. Later, other generalizations of spherical space-times such as solutions including perfect fluid or cosmological constant were also found \cite{Griffiths2009}.

Solutions including cylindrical symmetry were studied in a similar way. The vacuum cylindrical solution is called the Levi-Civita metric. This solution describes a massive infinitely long string or cylinder \cite{Griffiths2009}. Space-time around a charged massive string is called Raychaudhuri space-time \cite{Stephani2003}. A stationary (rotating) solution is described by the Lewis metric \cite{Herrera:2001xk,Mashhoon:1998fj,daSilva:2002zw}. A solution with a magnetic field $A_{z}\equiv A_{2}\neq 0$ was studied by Bonnor \cite{Stephani2003}. There are many other solutions such as for example the Majumdar-Papapetrou cylindrical metric \cite{Cermak:2012pt} describing space-time around massive charged cylinder where attracting gravitational forces are balanced by repulsive electrostatic forces.

In this article we will continue in the work initiated by the above mentioned authors. We will find stationary cylindrically-symmetric electro-vacuum solutions and compare them with known solutions.

\section{Einstein-Maxwell field equations in a general case}

The standard General Relativity with Einstein--Hilbert Lagrangian and geometry described by Levi--Civita connection, i.e. Riemann tensor and its contractions, are determined solely by the metric tensor $g_{\alpha\beta}$ and its derivatives. The equations of motion for the metric (Einstein equations) have following form
\begin{equation}
R_{\alpha\beta}-\frac{1}{2}g_{\alpha\beta}R\equiv
G_{\alpha\beta}=KT_{\alpha\beta}+\Lambda g_{\alpha\beta}\label{2}
.\end{equation}
where
$K$ is coupling constant
$K=\frac{8\pi\kappa}{c^{4}}$ (where $c$ is speed of light and $\kappa$ is Newton's gravitational constant),
$R_{\alpha\beta}$ is Ricci tensor, $R~\equiv~g^{\alpha\beta}R_{\alpha\beta}$ is the Ricci scalar, $G_{\alpha\beta}$ is Einstein tensor, $T_{\alpha\beta}$ is the total stress-energy-momentum tensor and $\Lambda$ is cosmological constant.

The stress-energy-momentum tensor of the electro--magnetic (EM) field is
\begin{equation}
T_{\alpha\beta}=\varepsilon_{0}c^{2}\left(\frac{1}{4}F^{\gamma\delta}F_{\gamma\delta}g_{\alpha\beta}
-F_{\alpha}\;^{\delta}F_{\beta\delta}\right),\label{3}
\end{equation}
where $\varepsilon_{0}$ is constant permittivity of vacuum
and $F_{\alpha\beta}$ is EM (strength) tensor.

The EM field is described by gauge vector potential $A_{\alpha}$ that enters equation~\eqref{3} through the EM (strength) tensor $F_{\alpha\beta}=A_{\beta,\alpha}-A_{\alpha,\beta}.$
The equation of motion for the EM field, the Maxwell equation, relates the strength tensor $F_{\alpha\beta}$ to electric current $j^{\alpha}$ as
\begin{equation}
{F}^{\alpha\beta}_{\;\;\;\; ;\alpha} = \frac{1}{\sqrt{|g|}}\left(
\sqrt{|g|}{F}^{\alpha\beta}\right)_{,\alpha} =
\frac{j_{\rm e}^{\beta}}{\varepsilon_{0}c^{2}},\label{4}
\end{equation}
where we have used identity relating covariant divergence of an anti-symmetric tensor with "ordinary" divergence.

\smallskip

In this article we will use the cylindrical coordinate system $x^{\alpha}\equiv(x^{0},x^{1},x^{2},x^{3})=(ct,r,\varphi,z)$. Derivative by radial coordinate $r$ we will note as $f_{,1}.$ Minkowski metric tensor will be taken with signature $\eta_{\alpha\beta}=\textup{diag}(1,-1,-1,-1).$

\section{Formulation of the problem}

In this work we will search for electro-vacuum stationary cylindrically-symmetric solutions. Einstein equations will be expressed with a non-zero cosmological constant. A natural choice of the EM tensor compatible with the assumed symmetry is $F_{01}=-F_{10}$, $F_{21}=-F_{12}$. We will not consider other components of the EM field.

The non-zero components correspond to radial electric field and magnetic field along $z$-axis respectively (in a flat space-time). A suitable ansatz for the gauge 4-potential $A_{\alpha}$ is   
$$A_{\alpha}=(A_{0},0,A_{2},0)\;\Rightarrow \;F_{01}=-A_{0,1}, \;\;F_{21}=-A_{2,1}.$$

\medskip

General form of cylindrical stationary metric tensor $g_{\alpha\beta}$ was found by Lewis and Papapetrou \cite{Stephani2003, Griffiths2009}. Weyl-Lewis-Papapetrou cylindrical stationary form can be written in the form
\begin{equation}
\mathrm{d} s^{2}=\frac{\left(c\mathrm{d}t+A\mathrm{d}\varphi\right)^{2}}{F}-BF\left(\mathrm{d} r^{2}+\mathrm{d}z^{2}\right)-WF\mathrm{d}\varphi^{2},
\label{1}\end{equation}
where $F$, $B$, $A$, $W$ are functions of coordinate $r\equiv x^{1}$.

\section{Equations of motion}
\subsection{Maxwell equations}
For any EM field with no charge or current source ($j_{\mathrm{e}}^{\alpha}=0$) Maxwell equation (Ampere and Gauss laws) can be written in the form $\left(\sqrt{|g|}{F}^{\alpha\beta}\right)_{,\alpha}=0$, or (with non-zero $F_{01}$, $F_{21}$) in the form 
\begin{equation*}
\left(\sqrt{|g|}{F}^{1\beta}\right)_{,1} =0\longrightarrow \sqrt{|g|}{F}_{1\gamma}g^{11}g^{\beta\gamma}=C^{\beta1}\longrightarrow 
\end{equation*}
\begin{equation}
\longrightarrow
C_{1}\sqrt{\left|W\right|}=F_{01}\left(\frac{A^{2}}{F}-FW\right)-F_{21}\frac{A}{F}, \;\;C_{2}\sqrt{\left|W\right|}=-\frac{1}{F}\left(F_{01}A-F_{21}\right),\longrightarrow \label{5rv}
\end{equation}
\begin{equation}
\longrightarrow
F_{01}=-\frac{C_{2}}{F\sqrt{\left|W\right|}}\left(A+\frac{C_{1}}{C_{2}}\right), \;\;F_{21}=-\frac{C_{2}A}{F\sqrt{\left|W\right|}}\left[\frac{C_{1}}{C_{2}}-\left(\frac{F^{2}W}{A}-A\right)\right],\label{5rvxy}
\end{equation}
where $C_{1}$ and $C_{2}$ are components of constants of integration $C^{\beta1}$ ($C^{01}=-C_{1}$, $C^{21}=-C_{2}$). In the text we will assume that 
$W$ can only be positive. The absolute value brackets in term $\sqrt{\left|W\right|}$ will not be used any more.

\subsection{Einstein equations}
The stress-energy-momentum tensor for an EM field with non-zero components of EM strength tensor $F_{01}$, $F_{21}$ and metric tensor given by \eqref{1} can be written in the form
\begin{eqnarray}
T_{00}&=&\frac{\varepsilon_{0}c^{2}}{2}\left(F^{\alpha1}F_{\alpha1}g_{00}-2F_{01}^{2}g^{11}\right),\\
T_{02}&=&\frac{\varepsilon_{0}c^{2}}{2}\left(F^{\alpha1}F_{\alpha1}g_{02}-2F_{01}F_{21}g^{11}\right),\\
T_{22}&=&\frac{\varepsilon_{0}c^{2}}{2}\left(F^{\alpha1}F_{\alpha1}g_{22}-2F_{21}^{2}g^{11}\right),\\
T_{11}&=&-\frac{\varepsilon_{0}c^{2}}{2}F^{\alpha1}F_{\alpha1}g_{11},\\
T_{33}&=&\frac{\varepsilon_{0}c^{2}}{2}F^{\alpha1}F_{\alpha1}g_{33}.
\label{6rv}
\end{eqnarray}

\medskip
Einstein equations contain a lot of terms in general. Therefore it is useful to work with linear combination of single equations. The combination $G_{11}g_{33}+G_{33}g_{11}=2g_{11}g_{33}\Lambda$ gives an equation describing dependence between cosmological constant and functions $F$, $B$, $W$ in the form
\begin{equation}
\sqrt{W}_{,11}=-2\Lambda\sqrt{W}FB\label{7rv}.
\end{equation}

Using the knowledge that the trace of an EM stress-energy-momentum tensor is equal to zero $T_{\alpha\beta}g^{\alpha\beta}=0$ in four-dimensional space-time we can obtain an equation containing only metric terms and cosmological constant. Using the equation \eqref{7rv} we obtain 
\begin{equation}
2\left(\frac{B_{,1}}{B}\right)_{,1}+2\left(\frac{F_{,1}}{F}\right)_{,1}+\frac{F_{,1}}{F}\left(\frac{F_{,1}}{F}+\frac{W_{,1}}{W}\right)=\frac{\left(A_{,1}\right)^{2}}{F^{2}W}\label{7,1rv}.
\end{equation}
The other four equations are combination of equations (for brevity we mention only the left hand side of equations) $G_{00}g_{02}-G_{02}g_{00}$, $G_{\alpha\beta}g^{\alpha\beta}g_{33}-2G_{33}$, $G_{00}g_{22}-G_{22}g_{00}$, $G_{02}g_{22}-G_{22}g_{02}$. All of these equations can be integrated (with help of Maxwell equations \eqref{5rv}). Integration decreases the order of derivative and naturally in equations emerge new constants of integration. After simple modifications we can write
\begin{equation}
\begin{array}{rcl}
\alpha C_{2}A_{0}&=&-K_{1}-\frac{A_{,1}}{2F^{2}\sqrt{W}},\\
\alpha C_{1}A_{2}&=&K_{4}-\sqrt{W}A\frac{F_{,1}}{F}-A\frac{W_{,1}}{2\sqrt{W}}+\frac{A^{2}A_{,1}}{2F^{2}\sqrt{W}}+\frac{A_{,1}\sqrt{W}}{2},\\
\alpha C_{1}A_{0}&=&-K_{2}-\sqrt{W}\left(\frac{F_{,1}}{F}+\frac{B_{,1}}{2B}\right)+\frac{AA_{,1}}{2F^{2}\sqrt{W}},\\
\alpha C_{2}A_{2}&=&-K_{3}+\frac{W_{,1}}{2\sqrt{W}}-\sqrt{W}\frac{B_{,1}}{2B}-\frac{AA_{,1}}{2F^{2}\sqrt{W}},\label{13rv}
\end{array}
\end{equation}
where $\alpha=\varepsilon_{0}c^{2}K.$ 

The system of equations \eqref{5rv}, \eqref{7rv}, \eqref{7,1rv}, \eqref{13rv} is complete system needed for searching  solutions with $\Lambda\neq0,$ $A\neq0,$ $A_{0}\neq0$, $A_{2}\neq0$. It is easy to find a particular solutions where $\Lambda=0$ together with $A=0$ or $A_{0}=0.$ These solutions can be easily found in general form and are presented in the next text.

The solution $\Lambda=0,$ $A\neq0,$ $A_{0}\neq0$ is more complicated to find in general form. We didn't find general form for this class of solutions. There is a subclass of solutions specified by the requirements $\Lambda=0,$ $A\neq0,$ $A_{0}\neq0$. For searching for this subsolutions we specified conditions for constants $C_{1}$ and $C_{2}.$  

\medskip

Consider $C_{1}\neq0$ and $C_{2}\neq0.$

Multiplying first and second equation of \eqref{13rv} we get equation
\begin{equation}
\left(\alpha C_{1}A_{2}-K_{4}\right)\left(\frac{1}{A}\right)_{,1}=-\left(\alpha C_{2}A_{0}+K_{1}\right)\left(\frac{F^{2}W}{A}-A\right)_{,1}\label{14rv}.
\end{equation}

In a similar spirit from \eqref{5rvxy} we obtain 

\begin{equation}
C_{1}A_{2,1}\left(\frac{1}{A}+\frac{C_{2}}{C_{1}}\right)=C_{2}A_{0,1}\left[\frac{C_{1}}{C_{2}}-\left(\frac{F^{2}W}{A}-A\right)\right]\label{15rv0}.
\end{equation}

The sum of \eqref{14rv} and \eqref{15rv0} can be integrated using the rule $(ab)'=a'b+ab'$
\begin{equation}
\left(\alpha C_{1}A_{2}-K_{4}\right)\left(\frac{1}{A}+\frac{C_{2}}{C_{1}}\right)=\left(\alpha C_{2}A_{0}+K_{1}\right)\left[\frac{C_{1}}{C_{2}}-\left(\frac{F^{2}W}{A}-A\right)\right]+K_{0}\label{15rv}.
\end{equation}

\smallskip

It is easy to find that suitable manipulation of the system of equations \eqref{13rv} lead to
\begin{equation}
\gamma_{1}C_{1}+\gamma_{2}C_{2}=\frac{C_{2}\sqrt{W}\left(C_{1}+C_{2}A\right)^{2}}{2F^{2}W}\left[\frac{F^{2}W}{C_{1}+C_{2}A}-\frac{A}{C_{2}}\right]_{,1}
,\label{18rv}
\end{equation}
\begin{equation}
\frac{\left[\frac{B}{W}\left(C_{1}+C_{2}A\right)\right]_{,1}}{\left[\frac{B}{W}\left(C_{1}+C_{2}A\right)\right]}=\frac{2\gamma_{1}}{C_{2}\sqrt{W}}-\frac{2}{C_{2}\sqrt{W}}\frac{\left(\gamma_{1}C_{1}+\gamma_{2}C_{2}\right)}{\left(C_{1}+C_{2}A\right)},\label{19rv}
\end{equation}
where constants $\gamma_{1},$ $\gamma_{2}$ are $\gamma_{1}=K_{1}C_{1}-K_{2}C_{2},$ $\gamma_{2}=K_{3}C_{1}+K_{4}C_{2}$.

It can be find that $\gamma_{1}C_{1}+\gamma_{2}C_{2}=-C_{1}C_{2}K_{0}.$

\medskip
 
\section{Exact solutions to Einstein equations with $\Lambda=0$}

In the previous section we have obtained a system of equations \eqref{5rv}, \eqref{7rv}, \eqref{7,1rv}, \eqref{13rv}, which will be used to determine the metric functions. It is difficult to find exact solution to the above mentioned system. It is useful to impose simplifying assumptions. First we are interested in the case of zero cosmological constant $\Lambda=0.$ This assumption simplifies equations \eqref{7rv} which can be easily integrated into the form $W=r^{2},$ where the constant of integration disappears after coordinate transformation. Yet, it is still hard to find a general solution to equations \eqref{5rvxy}, \eqref{7,1rv}, \eqref{13rv}.

The second simplification is achieved using the condition $\gamma_{1}C_{1}+\gamma_{2}C_{2}=0 \Leftrightarrow K_{0}=0.$ 
This enables us to write equations \eqref{18rv} and \eqref{19rv} in form
$$B=C_{3}r^{2\left(K_{1}\frac{C_{1}}{C_{2}}-K_{2}+1\right)},\;\; F^{2}r^{2}=\left(A-k_{0}\frac{C_{2}}{C_{1}}\right)\left(A+\frac{C_{1}}{C_{2}}\right).$$ 
Hence, the Einstein equation \eqref{7,1rv} becomes (see \cite{thesis})
$$\left[\left(\frac{Fr}{A+\frac{C_{2}}{C_{1}}}\right)^{\frac{1}{2}}\right]_{,11}=\frac{\frac{3}{4}+K_{1}\frac{C_{1}}{C_{2}}-K_{2}}{r^{2}}\left[\left(\frac{Fr}{A+\frac{C_{2}}{C_{1}}}\right)^{\frac{1}{2}}\right].$$

The solution to the above equations can be written in form
 \begin{equation}
BF=f^{2}r^{2a^{2}},
\;F=\frac{C}{\alpha_{1}r^{2}f^{2}+\frac{\alpha_{2}}{f^{2}}},
\;A=\beta_{1}\frac{F}{f^{2}}+\beta_{2},
\label{19rvx}
\end{equation}
where $\alpha_{1},$ $\alpha_{2},$ $\beta_{1},$ $\beta_{2},$ $a,$ $C$ are constants. $f$ is a function of the radial coordinate and will be explained in the next text. Advantage of this solution is that it can be used not only for case $A\neq0,$ $A_{0}\neq0.$ 

Maxwell equations \eqref{5rvxy} and Einstein equations \eqref{7,1rv}, \eqref{13rv} can be written as   
\begin{equation}
\begin{array}{rcl}
F_{01}&=&-\frac{C_{2}}{r}\left\lbrace\frac{1}{f^{2}}\left[\beta_{1}+\frac{\alpha_{2}}{C}\left(\beta_{2}+\frac{C_{1}}{C_{2}}\right)\right]+r^{2}f^{2}\frac{\alpha_{1}}{C}\left(\beta_{2}+\frac{C_{1}}{C_{2}}\right)\right\rbrace,\\ 
F_{21}&=&-\frac{C_{2}}{r}\left\lbrace\frac{\beta_{1}\beta_{2}}{f^{2}}-\frac{C}{\alpha_{1}f^{2}}\frac{\alpha_{1}r^{2}f^{2}-\frac{\alpha_{1}\beta_{1}^{2}}{f^{2}}}{\alpha_{1}r^{2}f^{2}+\frac{\alpha_{2}}{f^{2}}}+\frac{A}{F}\left(\beta_{2}+\frac{C_{1}}{C_{2}}\right)\right\rbrace,\\
\frac{f_{,11}}{f}+\frac{f_{,1}}{fr}-\frac{a^{2}}{r^{2}}&=&\left(\alpha_{1}\beta_{1}^{2}+\alpha_{2}\right)\left(1+\frac{2f_{,1}}{f}r\right)\frac{r\alpha_{1}}{\alpha_{1}r^{2}f^{2}+\frac{\alpha_{2}}{f^{2}}},\\ 
\alpha C_{2}A_{0}+K_{1}&=&\frac{\alpha_{1}\beta_{1}}{C}\left(1+\frac{2f_{,1}}{f}r\right) ,\\
\alpha C_{1}A_{2}-K_{4}&=&\left(1+\frac{2f_{,1}}{f}r\right)\left[ \beta_{2}\left(\frac{\alpha_{1}r^{2}f^{2}-\frac{\alpha_{1}\beta_{1}^{2}}{f^{2}}}{\alpha_{1}r^{2}f^{2}+\frac{\alpha_{2}}{f^{2}}}-\frac{\alpha_{1}\beta_{1}\beta_{2}}{C}\right)-\frac{A}{f^{2}}\frac{\alpha_{1}\beta_{1}^{2}+\alpha_{2}}{\alpha_{1}r^{2}f^{2}+\frac{\alpha_{2}}{f^{2}}}\right],\\
\alpha C_{1}A_{0}+K_{2}&=&1-a^{2}-\left(\frac{\alpha_{1}\beta_{1}^{2}+\alpha_{2}}{\alpha_{1}r^{2}f^{2}+\frac{\alpha_{2}}{f^{2}}}\frac{1}{f^{2}}+\frac{\alpha_{1}\beta_{1}\beta_{2}}{C}\right) \left(1+2r\frac{f_{,1}}{f}\right),\\
\alpha C_{2}A_{2}+K_{3}&=&1-a^{2}+\left(1+\frac{2f_{,1}}{f}r\right)\left(\frac{\alpha_{1}\beta_{1}\beta_{2}}{C}-\frac{\alpha_{1}r^{2}f^{2}-\frac{\alpha_{1}\beta_{1}^{2}}{f^{2}}}{\alpha_{1}r^{2}f^{2}+\frac{\alpha_{2}}{f^{2}}}\right).
\end{array}\label{1h}
\end{equation}

Potentials $A_{\alpha}$ and $A_{\alpha}+q_{,\alpha}$ (where $q$ is an arbitrary calibration function) are equivalent. Therefore we can use this freedom for recalibration of $A_{\alpha}.$ Without loss on generality we can shift potential $\alpha C_{2}A_{0}+K_{1}\rightarrow\alpha C_{2}A_{0}$ and $\alpha C_{1}A_{2}-K_{4}\rightarrow\alpha C_{1}A_{2}.$

For simplification in equations \eqref{1h} the condition $1-a^{2}=K_{2}-\frac{C_{1}}{C_{2}}K_{1}=K_{3}+\frac{C_{2}}{C_{1}}K_{4}$ can be used. The right hand side of this equation can be written in the form $\gamma_{1}C_{1}+\gamma_{2}C_{2}=0$.

\medskip

A solution of the function $f$ can be found in the form  
\begin{equation}
\begin{array}{rcl}
f&=&\left(k_{1}r^{a}+k_{2}r^{-a}\right), \;\;\; a^{2}>0, \\
f&=&k_{1}+k_{2}\ln\left(r\right)=k_{2}\ln\left(\frac{r}{r_{0}}\right),
\;\;\; a^{2}= 0.
\end{array}
\label{22rva}
\end{equation} 
Equations \eqref{22rva} describes solution $a^{2}\geq0$. If we use substitution $a=i\tilde{a}$, where $i$ is an imaginary unit ($i^{2}=-1$), we can find a new solution with condition $a^{2}<0$. Lets assume new constants $\tilde{k}_{1}=k_{1}+k_{2}$, $\tilde{k}_{2}=k_{1}-k_{2}$.
Then the first equation of \eqref{22rva} can be written as
\begin{equation*}
\begin{array}{rcl}
f&=&k_{1}r^{i\tilde{a}}+k_{2}r^{-i\tilde{a}}\\&=&\frac{\tilde{k}_{1}}{2}\left(r^{i\tilde{a}}+r^{-i\tilde{a}}\right)+\frac{\tilde{k}_{2}}{2}\left(r^{i\tilde{a}}-r^{-i\tilde{a}}\right)=
 \\
&=&\frac{\tilde{k}_{1}}{2}\left(e^{i\tilde{a}\ln\left(r\right)}+e^{-i\tilde{a}\ln\left(r\right)}\right)+\frac{\tilde{k}_{2}}{2}\left(e^{i\tilde{a}\ln\left(r\right)}-e^{-i\tilde{a}\ln\left(r\right)}\right)=
\\
&=&\tilde{k}_{1}\cos\left[\tilde{a}\ln\left(r\right)\right]+i\tilde{k}_{2}\sin\left[\tilde{a}\ln\left(r\right)\right]=
k\cos\left[\tilde{a}\ln\left(\frac{r}{r_{0}}\right)\right].
\label{nevim3}
\end{array}
\end{equation*}
Waves above $k_{1}$, $k_{2}$ will not be written in the next text. The imaginary unit will be included in constant $k_{2}$. The new solution (with $a^{2}<0$) is
\begin{equation}
f=k_{1}\cos\left[\tilde{a}\ln\left(r\right)\right]+k_{2}\sin\left[\tilde{a}\ln\left(r\right)\right], \;\; -\tilde{a}^2=a^2<0.
\label{22rv}
\end{equation}

\medskip

If we use solutions of function $f$ \eqref{22rva}, \eqref{22rv} we get zero on the left hand side of the third equation of the system \eqref{1h}. There are three different possibilities to get zero on the right hand side of the third equation of \eqref{1h}: 
\begin{itemize}
\item $\alpha_{1}=0,$
\item $\alpha_{1}\beta_{1}^{2}+\alpha_{2}=0,$
\item $1+\frac{2f_{,1}}{f}r=0.$
\end{itemize}
The third condition gives a trivial solution for the function $f$, which is a sub-solution of \eqref{22rva} without the EM field. In the next text we will not be interested in this condition. The first two conditions can be written in united form $\alpha_{1}+d\frac{\alpha_{2}}{\beta_{1}^{2}}=0,$ where $d=1,$ or $d=0.$

\subsection{Stationary electro-vacuum solutions}

Consider $A\neq0$ and $A_{0}\neq0$. As we said before we didn't find general solution in this case. Let us consider $\alpha_{1}\beta_{1}^{2}+\alpha_{2}=0$ (or $d=1$) and $\gamma_{1}C_{1}+\gamma_{2}C_{2}=0$, where $C_{1}\neq0$ and $C_{2}\neq0.$ We can see that fourth and sixth (and fifth and seventh) equations of system \eqref{1h} are identical if $\beta_{2}=-\frac{C_{1}}{C_{2}}.$ Assuming it hold, the system of equations \eqref{1h} reduces to  
\begin{equation}
\begin{array}{rcl}
F_{01}&=&-\frac{C_{2}\beta_{1}}{f^{2}r},\\ 
F_{21}&=&-\frac{C_{2}}{f^{2}r}\left(\beta_{1}\beta_{2}-\frac{C}{\alpha_{1}}\right),\\
\alpha C_{2}A_{0}&=&\frac{\alpha_{1}\beta_{1}}{C}\left(1+\frac{2f_{,1}}{f}r\right) ,\\
\alpha C_{2}A_{2}&=&\left(1+\frac{2f_{,1}}{f}r\right)\left(\frac{\alpha_{1}\beta_{1}\beta_{2}}{C}-1\right).\label{1hx}
\end{array}
\end{equation}

By derivative of the second pair of the system \eqref{1hx} and using $F_{\alpha1}=-A_{\alpha,1}$ we get conditions for constants $k_{1}$ and $k_{2}$  
\begin{equation}
\left(1+\frac{2f_{,1}}{f}r\right)_{,1} =
\left\{\begin{array}{ll}
\medskip
8k_{1}k_{2}\frac{a^{2}}{rf^{2}},\; &a^{2} > 0,\\
\medskip
-2k_{2}^{2}\frac{1}{rf^{2}},\; &a^{2} = 0,\\
2\left(k_{1}^{2}+k_{2}^{2}\right)\frac{a^{2}}{rf^{2}},\; &a^{2} < 0,\\
\end{array}\right.
\label{abbreviate.k}\end{equation}
i.e.
\begin{equation}
\begin{array}{rcl}
\medskip
8k_{1}k_{2}a^{2}&=\frac{C}{\alpha_{1}}C_{2}^{2}\alpha,\; &a^{2} > 0,\\
\medskip
-2k_{2}^{2}&=\frac{C}{\alpha_{1}}C_{2}^{2}\alpha,\; &a^{2} = 0,\\
2\left(k_{1}^{2}+k_{2}^{2}\right)a^{2}&=\frac{C}{\alpha_{1}}C_{2}^{2}\alpha,\; &a^{2} < 0.
\end{array}
\label{abbreviate.ka}\end{equation}

\begin{itemize}
\item $A\neq0$, $A_{0}\neq0$, $A_{2}\neq0$: From the system of equations \eqref{1hx} we can see that non-zero components of four-potential are linearly proportional. We can write this dependence in the form
$$A_{0}k_{0}\frac{C_{2}}{C_{1}}=A_{2},$$
where $k_{0}\frac{C_{2}}{C_{1}}$ is constant of proportionality. From the system of equations \eqref{1hx} we can see that $C=-\alpha_{1}\beta_{1}(\frac{C_{1}}{C_{2}}+k_{0}\frac{C_{2}}{C_{1}}).$ Using of coordinate transformation we can put $\alpha_{1}=\beta_{1}=1\rightarrow\alpha_{2}=-1.$ 
\item $A\neq0$, $A_{0}\neq0$, $A_{2}=0$: Assume the same results as in the previous case. We can get a solution with zero $A_{2}$ if constant $k_{0}=0.$ In the same way as in the previous case we can write $C=-\alpha_{1}\beta_{1}\frac{C_{1}}{C_{2}}$ and $\alpha_{1}=\beta_{1}=1\rightarrow\alpha_{2}=-1.$ 
\end{itemize}

\subsection{Solution with no electric field}

\begin{itemize}
\item $A\neq0$, $A_{0}=0$, $A_{2}\neq0$: From Maxwell equations \eqref{5rv} we can see that $F_{01}=0$ together with $F_{21}\neq0$ can be satisfied only if $A=\beta_{2}=-\frac{C_{1}}{C_{2}}$. Using coordinate transformation we get $A=0$ that describes the static solution. In the next text we will show that the solution with $\alpha_{1}=0$ also give a static solution i.e. $A=0$. Therefore there is no solution $A\neq0$, $A_{0}=0$, $A_{2}\neq0.$ 
\item $A=0$, $A_{0}=0$, $A_{2}\neq0$: As we mentioned in the previous case for $A_{0}=0$ together with $A_{2}\neq0$ the equation $\beta_{1}=0\rightarrow \alpha_{2}=0$ must hold (from condition $\alpha_{1}\beta_{1}^{2}+\alpha_{2}=0$). With the help of coordinate transformation we can find $C=\alpha_{1}=1.$ A change of signatures of constants $C$ or $\alpha_{1}$ change the signature of metric. This (by contrast to previous solutions) reduces the solution to case $a^{2}>0,$ because equations \eqref{abbreviate.ka} for $a^{2}=-\tilde{a}^{2}\leq0$ can not be satisfied. This solution was found by Bonnor \cite{Stephani2003,thesis}.
\end{itemize}

\subsection{Vacuum stationary solution}\label{kap}

\begin{itemize}
\item$A\neq0$, $A_{0}=0,$ $A_{2}=0:$ This solution is quite different from previous solutions. We again consider solution in the form \eqref{19rvx}. The first two equations of \eqref{1h} (Maxwell equations) in the case of $A_{0}=A_{2}=0\rightarrow C_{1}=C_{2}=0$ are identically equal to zero. Zero EM field gives a constant on the left hand side of the fourth, fifth, sixth and seventh equations of \eqref{1h}. With the condition $\alpha_{1}\beta_{1}^{2}+\alpha_{2}=0$ we can find solution to the system of equations \eqref{1h} in the form $f=k_{1}r^{a}.$ This solution corresponds to the Lewis solution \cite{Herrera:2001xk,Mashhoon:1998fj,daSilva:2002zw}.
\end{itemize}

\medskip

We can write this solution to the last point in the form
\begin{equation}
\begin{array}{rcl}
Fr&=&\frac{C}{\alpha_{1}k_{1}^{2}r^{1+2a}+\frac{\alpha_{2}}{k_{1}^{2}r^{1+2a}}}=\frac{C}{\alpha_{1}f'+\frac{\alpha_{2}}{f'}}, 
\\ 
BF&=&k_{1}^{2}r^{2a^{2}+2a}=k_{1}^{2}r^{\frac{b^{2}-1}{2}},
\\
A&=&\beta_{1}\frac{Fr}{f'}+\beta_{2},
\label{aba}
\end{array}
\end{equation}
where $f'=k_{1}^{2}r^{b},$ $b=1+2a.$ A combination of solution \eqref{aba} with equations \eqref{7,1rv} and \eqref{13rv} gives conditions for the constants in the form 
$$b=K_{2}-K_{3}+2K_{1}\beta_{2},\;\;\beta_{2}b=\beta_{2}^{2}K_{1}-K_{4},\;\;b^{2}=3-2\left(K_{2}+K_{3}\right) ,\;\;\alpha_{1}b=-K_{1}\beta_{1}C .$$
It is obvious for this solution that the condition $b^{2}>0.$ is satisfied. Similarly as in previous cases we can find solutions for $b^{2}\leq0.$

\medskip

The new solution corresponding to $b=0\rightarrow a=-\frac{1}{2}$ can be written in the form
\begin{equation}
Fr=\frac{1}{2K_{1}\ln\left(\frac{r}{r_{0}}\right)},\;\;BF=k_{1}^{2}r^{-\frac{1}{2}}
,\;\;A=Fr+\beta_{2}.\label{aab}
\end{equation}
Solution \eqref{aab} is again a solution to the system of equations \eqref{7,1rv} and \eqref{13rv} with zero EM field. Equations \eqref{13rv} gives conditions for constants: $K_{2}+K_{3}=\frac{3}{2},$ $K_{4}=K_{1}\beta_{2}^{2},$ $K_{2}-K_{3}=-2K_{1}\beta_{2}.$ This solution corresponds to the Datta-Raychaudhuri solution \cite{Gersl:2003vr}.
With the help of coordinate transformation we can write $\beta_{2}=0.$

\medskip

A solution corresponding to $b^{2}=-\tilde{b}^{2}<0$ i.e. $b=i\tilde{b}$ can be written in the form
\begin{equation}
Fr=\frac{C}{\cos\left[ \tilde{b}\ln\left(\frac{r}{r_{0}} \right)\right]},\;\;A=\pm Fr \sin\left[ \tilde{b}\ln\left(\frac{r}{r_{0}} \right)\right]+\beta_{2},\;\;BF=k_{1}^{2}r^{-\frac{\tilde{b}^{2}+1}{2}}.
\end{equation} 
This solution again corresponds to a solution of the system of equations \eqref{7,1rv} and \eqref{13rv} with zero EM field. The system of equations gives conditions for constants in the form $\tilde{b}=\mp 2CK_{1}$, $K_{4}=K_{1}\left(\beta_{2}^{2}+C^{2}\right),$ $2K_{1}\beta_{2}=K_{3}-K_{2},$ $1-K_{2}-K_{3}=-\frac{\tilde{b}^{2}+1}{2}.$

\subsection{Static solutions}
Static solutions are given by the condition $A=0.$ With this condition equations \eqref{13rv} can be reduced to
\begin{equation}
\begin{array}{rcl}
\alpha C_{2}A_{0}&=&-K_{1},\\
\alpha C_{1}A_{2}&=&K_{4},\\
\alpha C_{1}A_{0}&=&-K_{2}-r\left(\frac{F_{,1}}{F}+\frac{B_{,1}}{2B}\right),\\
\alpha C_{2}A_{2}&=&-K_{3}+1-r\frac{B_{,1}}{2B}.\label{13rvxd}
\end{array}
\end{equation}

\begin{itemize}
\item $A=0$, $A_{0}\neq0$, $A_{2}\neq0:$ There is no reason to consider $A_{\alpha}$ potential to be a non-zero constant. Therefore there are two possibilities $C_{1}=A_{0}=0$ or $C_{2}=A_{2}=0.$ This implies that there are no static electro-vacuum cylindrical solutions with non-zero $A_{0}$ together with non-zero $A_{2}.$  
\end{itemize}

The solution with $A_{0}=0$ was studied in the previous section. It is easy to prove that the solution for $C_{2}=A_{2}=0$ can be written in the form which is a reduced solution \eqref{19rvx} where $\alpha_{1}=0$
\begin{equation}
B=r^{2a^{2}},\;\;F=f^{2},\;\;A=\beta_{1}+\beta_{2},\label{cba}
\end{equation}
where we used coordinate and constant transformation for $C=\alpha_{2}=1.$ From solution \eqref{cba} it is evident that $A$ is a constant and can be easily transformed into zero. Using this solution in \eqref{1h} we obtain non-trivial equations (of \eqref{1h}) in the form
\begin{equation}
\begin{array}{rcl}
F_{01}&=&-\frac{C_{1}}{f^{2}r},\\
0&=&-\frac{a^{2}}{r^{2}}+\frac{f_{,11}}{f}+\frac{f_{,1}}{f}\frac{1}{r},\\
\alpha C_{1}A_{0}&=&-K_{2}-2r\frac{f_{,1}}{f}-a^{2}.\\
\label{cba1}\end{array}
\end{equation}

\begin{itemize}
\item $A=0$, $A_{0}\neq0$, $A_{2}=0:$ For non-zero $A_{0}$ the condition $C_{1}\neq0$ must hold (Functions $B,$ $F$ are given by \eqref{cba}). The solution for function $f$ can be found in same form as in previous cases (\eqref{22rva} or \eqref{22rv}). As in the previous section, from the last equation and the first equation of \eqref{cba1}, we obtain conditions for constants $k_{1}$ and $k_{2}$ in the form
\begin{equation}
\begin{array}{rcl}
\medskip
-8k_{1}k_{2}a^{2}&=C_{1}^{2}\alpha,\; &a^{2} > 0,\\
\medskip
2k_{2}^{2}&=C_{1}^{2}\alpha,\; &a^{2} = 0,\\
-2\left(k_{1}^{2}+k_{2}^{2}\right)a^{2}&=C_{1}^{2}\alpha,\; &a^{2} < 0.
\end{array}
\label{abbreviate.kb}\end{equation}
It can be seen that every one of these conditions can be satisfied. Solutions with $a^{2}>0$ and $a^{2}=0$ have been studied before by Raychaudhuri \cite{Stephani2003} and Majumdar and Papapetrou \cite{Cermak:2012pt,Gurses:1998zu,Lemos:2005md}. 
\item $A=0$, $A_{0}=0$, $A_{2}=0:$
The special case from the previous solution can be found when $k_{2}=0$ (or $k_{1}=0$). In this case the last equation of the system \eqref{cba1} is a constant on the right hand side. This leads to a zero EM field. The solution with $a^{2}>0$ ($f=k_{1}r^{a}$) is a famous cylindrical Levi-Civita solution \cite{Delice:2004wk,Griffiths2009} and solution $a^{2}=0$ i.e. ($f=k_{1}$) gives a flat Minkowski metric \cite{Griffiths2009}. It was found that in the case of Levi-Civita solution $a$ is connected (but not directly dependent) to the length mass density \cite{Griffiths2009}. The solution with $a^{2} < 0$ is not physical solution of Einstein equations.
\end{itemize}

\subsection{General Datta and Raychaudhuri and Islam, Bergh, Wils solutions}
We found a solution for $A\neq0$, $A_{0}\neq0$ with help of simplifying conditions $k_{0}=0$. As we have mentioned before there are other simplifying conditions that allow us to find a solution for $A\neq0$, $A_{0}\neq0$. 

\medskip
Consider $A_{2}=0$, $C_{1}=0,$  $C_{2}\neq0.$ If we use these conditions, equations \eqref{13rv} and Maxwell equations \eqref{5rv} reduce to the form 
\begin{equation}
\begin{array}{rcl}
\alpha C_{2}A_{0}&=&-K_{1}-\frac{A_{,1}}{2F^{2}r},\\
0&=&K_{4}-rA\frac{F_{,1}}{F}-A+\frac{A^{2}A_{,1}}{2F^{2}r}+\frac{A_{,1}r}{2},\\
0&=&-K_{2}-r\left(\frac{F_{,1}}{F}+\frac{B_{,1}}{2B}\right)+\frac{AA_{,1}}{2F^{2}r},\\
0&=&-K_{3}+1-r\frac{B_{,1}}{2B}-\frac{AA_{,1}}{2F^{2}r},\\
0&=&C_{2}F_{01}\left(F^{2}r^{2}-A^{2}\right),\\ 
C_{2}r&=&-\frac{A}{F}F_{01}.\label{30rv2}
\end{array} 
\end{equation}

 The solution to the system of equations \eqref{30rv2} and equation \eqref{7,1rv} can be written in the form
\begin{equation}
BF=r^{-\frac{1}{2}},\;\;A=Fr=\frac{1}{k_{1}+2\alpha C_{2}^{2}r+\left(2K_{1}+2\alpha C_{2}k_{2} \right)\ln \left(r\right)},\;\;A_{0}=C_{2}r+k_{2},
\end{equation}
provided that
$$K_{4}=0,\;\;K_{2}=K_{3}=\frac{3}{4},$$
where $k_{1},$ $k_{2}$ are constants of integration.

This solution is equivalent to the Datta and Raychaudhuri solution \cite{Gersl:2003vr}, where $k_{1}=0$ and $2\alpha C_{2}^{2}=4$. A special case of $\alpha C_{2}=0, k_{2}=0$ gives the solution mentioned in section \ref{kap}. 

\medskip
Similarly we can find solution of the system of equations \eqref{5rv}, \eqref{7rv}, \eqref{7,1rv}, \eqref{13rv} if we impose $C_{2}=0.$ Such a solution was founded by Islam, Bergh and Wils \cite{Stephani2003} and can it be written in the form
$$\frac{A}{a}=\frac{1}{F}=r^{\frac{2}{3}}, \;\; B=r^{\frac{2}{9}}e^{-a^{2}r^{\frac{2}{3}}}, \;\; A_{0}=\frac{3}{2}C_{1}a^{2}r^{\frac{2}{3}}, \;\;A_{0}=\frac{3}{4}C_{1}a^{3}r^{\frac{4}{3}}.$$

\section{Summary}

In this article we studied cylindrical solutions with an EM field. The different classes of solutions correspond mainly to the three parameters: function $A$ which describes stationarity of space-time and two components of EM four-potential $A_{0}$ and $A_{2}.$ The description of EM field is quite complicated in a non-static Einstein gravity field. Especially in the case of rotating space-times there can be components of four-potential which is difficult to interpret. For a better orientation we note that in flat space-time (cylindrical coordinate system) component $A_{0}$ describes a radial electric field and $A_{2}$ describes a magnetic field in $z$ direction.

We found that all solutions (except general solution of Datta-Raychaudhuri and Islam, Bergh, Wils solution) can be written in one general form  
$$\mathrm{d} s^{2}=\frac{\left(c\mathrm{d}t+A\mathrm{d}\varphi\right)^{2}}{F}-BF\left(\mathrm{d} r^{2}+\mathrm{d}z^{2}\right)-r^{2}F\mathrm{d}\varphi^{2},$$     
$$
BF=f^{2}r^{2a^{2}},
\;F=\frac{C}{\alpha_{1}r^{2}f^{2}+\frac{\alpha_{2}}{f^{2}}},
\;A=\beta_{1}\frac{F}{f^{2}}+\beta_{2},
$$
where function $f$ can be given in three different forms 
\begin{equation*}
\begin{array}{rcl}
f&=&\left(k_{1}r^{a}+k_{2}r^{-a}\right), \;\;\; a^{2}>0, \\
f&=&k_{1}+k_{2}\ln\left(r\right)=k_{2}\ln\left(\frac{r}{r_{0}}\right),
\;\;\; a^{2}= 0,\\
f&=&k_{1}\cos\left[\tilde{a}\ln\left(r\right)\right]+k_{2}\sin\left[\tilde{a}\ln\left(r\right)\right]=k\cos\left[\tilde{a}\ln\left(\frac{r}{r_{0}}\right)\right], \;\; -\tilde{a}^2=a^2<0.
\end{array}
\end{equation*} 
The third equation is equivalent to the first if $a=i\tilde{a}.$ This substitution can not be used in the case $A\neq0,$ $A_{\alpha}=0$. Here function $r^{2a+1}$ is replaced by $r^{b}$, where $b=2a+1$ and the operation can be realized with $\frac{C}{Fr}=\alpha_{1}r^{b}+\alpha_{2}r^{-b}$ (instead of $f$). There are two classes of solutions in dependence on $d$ parameter in the equation $\alpha_{1}+d\frac{\alpha_{2}}{\beta_{1}^{2}}=0,$ The first class fulfils $d=1$ and the second $d=0.$ We found that the general form of the solution can not be used for every case of cylindrical electro-vacuum solutions. Some solutions are forbidden.

All solutions (except general Datta-Raychaudhuri and Islam, Bergh, Wils solutions) are summed up in table \ref{tab1r}.

\begin{center}
\begin{table}[h]
\begin{center}
\begin{tabular}{|c|c|c|c|c|c|}
\hline
\multicolumn{6}{|c|}{\textbf{Cylindrical electro-vacuum stationary and static solutions}} 
\\
\hline \hline Type & \multicolumn{2}{c}{Conditions}  &  \multicolumn{3}{|c|}{Existence solution with parameter $a$ $/$ Name} 
\\
\hline 
 $A$,\; $A_{0}$,\; $A_{2}$ &$d\neq0$ &$k_{1}k_{2}\neq0$&
$a^{2}>0$ & $a^{2}=0$ &$a^{2}<0$
\\
 \hline $\neq0$, $\neq0$, $\neq0$ & \ding{51} & \ding{51} & \ding{51}&
\ding{51} & \ding{51}
\\
 \hline $\neq0$, $\neq0$, $=0$ & \ding{51} &  \ding{51} & \ding{51}&
\ding{51} & \ding{51}
\\
 \hline $\neq0$, $=0$, $\neq0$ & -- & -- & \ding{53}&
\ding{53} & \ding{53}
\\
 \hline $=0$, $=0$, $\neq0$ & \ding{51} & \ding{51} & \ding{51}$/$Bonnor&
\ding{53} & \ding{53}
\\
 \hline $\neq0$, $=0$, $=0$ & \ding{51} & \ding{53} & \ding{51}*$/$Lewis&
\ding{51}*$/$Datta-Raychaudhuri & \ding{51}*
\\
 \hline $=0$, $\neq0$, $\neq0$ & -- & -- & \ding{53}&
\ding{53} & \ding{53}
\\
 \hline $=0$, $\neq0$, $=0$ & \ding{53} & \ding{51} & \ding{51}$/$Raychaudhuri&
\ding{51}$/$Majudar-Papapetrou & \ding{51}
\\
 \hline $=0$, $=0$, $=0$ & \ding{53} & \ding{53} & \ding{51}$/$Levi-Civita&
\ding{51}$/$Minkowski & \ding{53}
\\
\hline 
\end{tabular}
\end{center}
\caption{* In the case of $A\neq0$, $A_{0}=0$, $A_{2}=0$ we are interested in three different solutions with $b^{2}>0,$ $b^{2}=0,$ $b^{2}<0,$ i.e. in this case $a$ is replaced by $b=2a+1$.}\label{tab1r}\end{table}
\end{center}
\normalsize

\section{Conclusion}

In this article we studied electro-vacuum stationary and static cylindrical solutions in general relativity. We were searching for solutions to Einstein-Maxwell equations in four dimensions (three space dimensions and one time dimension). The metric tensor includes three functions of radial coordinates $F, B, A$. The solution is described by three important parameters. First is function $A$ which describes stationarity of the solution. Second is the component of four-potential $A_{0}$, which is (in flat space-time case) responsible for the radial electric field. The third component is the component of four-potential $A_{2}$, which is (in flat space-time case) responsible for the magnetic field in $z$ direction. The same description of four potential can not be used in the case of general relativity. Solutions are divided according to dependence if those parameters are zero or not. Classification distinguishes eight different solutions. We found that two of them ($A=0$, $A_{0}\neq0$, $A_{2}\neq0$ and $A\neq0$, $A_{0}=0$, $A_{2}\neq0$) are forbidden.

We found a general form of solutions including all possible cases of combinations $A, A_{0}, A_{2}.$ This form includes known electro-vacuum cylindrical solutions except the general form of the Datta-Raychaudhuri and Islam, Bergh, Wils solutions. Solutions are dependent on parameter $a$ (or $b$ in case $A\neq0$, $A_{0}=0$, $A_{2}=0$). We distinguish three different cases of this parameter: $a$ is real $a^{2}>0$, $a$ is zero $a^{2}=0$, $a$ is imaginary $a^{2}<0$ (same for $b$ in case $A\neq0$, $A_{0}=0$, $A_{2}=0$). Not every solution for this parameter is allowed. Einstein-Maxwell equation implies conditions for constants of integration and some of the solutions break these conditions. All possible and forbidden cases are summed up in table \ref{tab1r}.

Our next research of the electro-vacuum cylindrical solutions should be focused on the generalization of mentioned solutions (in this work). For example adding a new non-zero component of four-potential $A_{3}$ (in flat space-time responsible for magnetic field in $\varphi$ direction).

The second focus of our next research following on from this work can be the exploration of class solutions with parameter $a$ with the condition $a^{2}=0$. It was proved in the case of the Majumdar-Papapetrou solution (one of the solution in the table \ref{tab1r} with $a^{2}=0$) that Einstein equations reduces to a linear differential equation (Poisson equation), where the metric components are directly dependent on field potential. It is a question of whether the same simplification can be found in the case of other solutions with $a^{2}=0.$ This assumption leads to a similarity between function $f$ and the potential of a cylindrical solution.

\section*{Acknowledgements}

Financial support by Department of Physical Electronics, Faculty of Science, Masaryk University is acknowledged.

Author is grateful to doc. RNDr. Vladim\'{i}r Balek, CSc. and to dear colleague Mgr. Martin Zouhar, Ph.D. for useful comments and suggestions to the manuscript.

\bibliographystyle{unsrt}
\bibliography{citace}

\end{document}